\documentclass[12pt]{article} 
\usepackage[tbtags]{amsmath}
\usepackage{amssymb}
\usepackage{amsfonts}
\usepackage{euscript}
\usepackage{graphicx}
\usepackage{color}

\newcommand{\mt}{\mathrm{Tr} }
\newcommand{\nn}{\nonumber }

\begin{document}

\title{ On the ring of local polynomial
           invariants for a pair of entangled
                                         qubits }

\author{V.~Gerdt\,${}^a$\,,\
        A.~Khvedelidze\,${}^{a,b}$\,, and \
        Yu.~Palii\,${}^{a,c}$
\\[0.7cm]
{${}^a$ } \small \it Laboratory of Information Technologies,\\
\small  \it Joint Institute for Nuclear Research, Dubna,
141980,  Russia
\\[0.5cm]
 ${}^b$ \small
\it Department of Theoretical Physics,  A. Razmadze Mathematical Institute,\\
\small  \it  Tbilisi GE-0193, Georgia
\\[0.5cm]
 {\small {${}^c$
\it Institute of Applied Physics, Chisinau MD-2028, Moldova}} }

\date{}

\maketitle

\begin{abstract}
The entanglement characteristics of two qubits are encoded  in the
invariants of the adjoint action of  $\
\mathrm{SU(2)}\otimes\mathrm{SU(2)}$ group on  the space of density
matrices $\mathfrak{P}_+\,,$ defined as the space of $4\times 4 $
positive semi-definite Hermitian matrices. The corresponding ring
$\mathbb{C}[\,\mathfrak{P}_+\,]^{\mathrm{SU(2)\otimes SU(2)}}\,$ of
polynomial invariants is studied.
The special integrity basis for the ring
$\mathbb{C}[\,\mathfrak{P}_+\,]^{\mathrm{SU(2)\otimes SU(2)}}\,$, is
described and constraints on its elements due to the positive
semi-definiteness of density matrices  are given explicitly in the
form of polynomial inequalities. The suggested  basis is
characterized by the property that only a minimal number of
invariants, namely two  primary invariants of degree 2, 3 and one
secondary invariant of degree 4 appearing in the Hironaka
decomposition of
$\mathbb{C}[\,\mathfrak{P}_+\,]^{\mathrm{SU(2)\otimes SU(2)}}\,,$
are subject to the polynomial inequalities.
\end{abstract}

\newpage

\section{Introduction}

According to the quantum theory non-locality of a quantum word
manifests itself in a way that is very different from the intuitive
classical views. At the very outset of quantum epoch reflections on
that fact create a variety of paradoxes beginning from the
Eistein\--Podolski\--Rosen paradox  and the famous neither dead nor
alive Schr\"{o}dinger cat~\cite{EPR,Bell,Aspect}. Only towards the
end of the XX-th century after the advances of technology, when a
manipulation with quantum coherency became reality, the pragmatic
approach to the problem rises the question of a practical usage of
quantum non-locality. Time for the realization of quantum
communications  and creation of the quantum computer has to
come~\cite{Nielsen}.

The difference between quantum and classical correlations has a very
transparent mathematical background. One can already see its roots
comparing the basic states of classical and quantum computers; bits
and qubits. While an arbitrary $n$-bit string can be transformed
into other by the so-called ``local transformation'',  acting on its
constituent  bits, in quantum case this is true for one qubit states
only. In other  words the action of ``local transformation'' cease
to be \textit{transitive} for multiqubit systems
\cite{Vedral,BengtssonZyczkowski}. The action of local
transformations splits the space of an arbitrary quantum system into
the equivalence classes in way that each class is characterized by a
different  non-local properties \cite{Werner}. Therefore the problem
of classification of non-localities in a system of $n$-qubits
reduces to the mathematical problem of description of orbits of
``local'' group action on the space of states
\cite{SchlienzMahler,LindenPopescu}. The corresponding orbit space,
$\mathcal{E}_n$, is termed as  \textit{``entanglement space''}
\cite{Vedral,BengtssonZyczkowski}. For its characterization the
mathematical formalism based on the classical theory of invariants
(c.f. \cite{Weyl,PopovVinberg}) is often applied. In this approach,
in order to separate orbits, i.e., to introduce coordinates on
$\mathcal{E}_n\,,$ the polynomials in elements of the density
matrices, which are invariant under the local transformation, are
used.

The entanglement space has highly nontrivial geometric  and
topological structure \cite{BengtssonZyczkowski,KusZyczkowski}.
Especially, a complexity of
$\mathcal{E}_n$ is steeply  rising with  number of qubits growing up.
This makes computations very
tedious. However, for the lowest, 2-qubits system, the approach
based on the classical theory of invariants allows to obtain a
series of important algebraic results, clarifying the properties of
$\mathcal{E}_2\,$
\cite{LindenPopescu,GrasslRottelerBeth,KingWelshJarvis}.

There is one further  complication  with the description  of
$\mathcal{E}_n$. In virtue of a physical requirement the density
matrices should be positive semi-definite
\cite{vonNeumann,Landau,Blum}. Therefore,  the space of local group
action is not a linear space, but represents a certain
semi-algebraic variety $\mathfrak{P}_+\,$. This
circumstance should be taken into account applying the classical
theory of invariants for construction of the  orbit space.
In the present article
this  problem is analyzed and detailed solution is given  for the case of
2-qubits.  With this aim  the semi-definiteness of density
matrices is formulated explicitly in the form of polynomial
inequalities in the scalars of the adjoint action of the group
${\mathrm{SU(2)\otimes SU(2)}}\,.$ Apart from this, the integrity
basis for the polynomial ring
$\mathbb{C}[\,\mathfrak{P}_+\,]^{\mathrm{SU(2)\otimes SU(2)}}\,$,
that includes the minimal number of elements subject to the above
inequalities, will be presented.

Our plan is as follows. We start, in paragraphs 2 and 3,  with the
brief review of necessary notions from quantum mechanics and put
them into context suitable for  the characterization of entanglement
within the classical theory of invariants. Further, in the 4-th
paragraph the system of polynomial inequalities in the Casimir
operators of enveloping algebra of $\mathrm{SU(4)}\,, $ that describes
the space $\mathfrak{P}_+\,$ is derived. Regarding to these
inequalities, in the last paragraph the integrity basis for the ring
$\mathbb{C}[\,\mathfrak{P}_+\,]^{\mathrm{SU(2)\otimes SU(2)}}\,$ is
constructed.

\section{The space of states }

A generic mixed state of $n$-level quantum system is described by
$n\times n$ complex matrix, the density matrix $\varrho\,$
\cite{vonNeumann,Landau}, satisfying following
conditions\footnote{The special class of idempotent matrices,  \(\
\varrho^2=\rho\,, \) corresponds to the so-called \textit{pure
states}, whose description reduces to the usage of rays in a Hilbert
space. A mixed state is a mixture of pure states.}:
\begin{enumerate}
  \item[i.]{\it  Hermicity } \---  $\varrho = \varrho^+\,,$
  \item[ii.] {\it finite trace } \---
$\mathrm{Tr}(\varrho) = 1\,,$
  \item[iii.] {\it positive semi-definiteness } \--- $\varrho \geq
  0$\,.
\end{enumerate}
Mixed states form the  subspace $\mathfrak{P}_+\,$, of the space of
Hermitian $n\times n$ matrices. It is instructive, before
considering a generic $n$-level system, to start  with the simplest
two-level quantum mechanical model.

\subsection{Qubit}

In quantum theory of information an abstract quantum mechanical
model with two levels takes a special place and independently of its
physical realization carries the universal name \--- \textit{qubit}.

The qubit state is given by a density matrix that coincides with the
standard density matrix of the non-relativistic spin 1/2:
\begin{equation}\label{eq:qbit}
\varrho =
\frac{1}{2}\,\left(1+\boldsymbol{\alpha}\cdot\boldsymbol{\sigma}\right)
\,,
\end{equation}
where $\boldsymbol{\sigma}$ \-- set of the Pauli matrices
\footnote{The explicit form of $\sigma$-matrices is given below, in
paragraph 5, formulaes  (\ref{PauliMatrices}).} and
$\boldsymbol{\alpha}$ is defined as mathematical expectation:
\[
\boldsymbol{\alpha}=\mathrm{Tr}\left(\boldsymbol{\sigma}\varrho \right)\,,
\]

In the representation (\ref{eq:qbit}) requirements  (i.) and (ii.)
are taken into account by construction. The condition (iii.)
restricts the parameter space of mixed states by a unit ball
\begin{equation}\label{eq:qball}
    \boldsymbol{\alpha}^2 \leq 1\,,
\end{equation}
while for the pure states of qubit  the expectation
$\boldsymbol{\alpha}$ lies on the  Bloch 2-sphere
\begin{equation} \nn 
\boldsymbol{\alpha}^2=1\,.
\end{equation}

\subsection{Qudit}

Analogously to the qubit the special terminology for  $d$ \-- level
quantum system,  a  \textit{``qudit''}, has been introduced. The
generalization of representation (\ref{eq:qbit})
to the case of qudits  reads~\cite{HioeEberly}:
\begin{equation}\label{eq:qdit}
\varrho = \frac{1}{d} \left(\mathbb{I}_d+\sqrt{\frac{d(d-1)}{2}}\,
\boldsymbol{\xi} \cdot \boldsymbol{\lambda}\right)\,,
\end{equation}
where  $\boldsymbol{\xi}= \langle
\boldsymbol{\lambda} \rangle \in \mathbb{R}^{d^2-1}\,$ is $d^2-1$- dimensional Bloch vector.
In the
expansion (\ref{eq:qdit}) components of the vector
$\boldsymbol{\lambda}=(\lambda_1, \lambda_2, \ldots,
\lambda_{d^2-1})$  represent the elements of the $\mathrm{su}({d})$
algebra normalized by  conditions
\[
\lambda_i\lambda_j=\frac{2}{d}\,\delta_{ij}\mathbb{I}_d+(d_{ijk}+i\,f_{ijk})\lambda_k,
\]
$\delta_{ij}$ is the Kronecker symbol. The constants $d_{ijk}$ and $f_{ijk}$ are the
so-called totally symmetric and
antisymmetric structure constants of  the algebra:
\[
d_{abc} = \frac{1}{4}\mbox{Tr}(\{\lambda_a,\lambda_b\}\lambda_c) ,
\qquad f_{abc} =
-\frac{i}{4}\mbox{Tr}([\lambda_a,\lambda_b]\lambda_c),
\]
where
\[\{\lambda_a,\lambda_b\}=\lambda_a \lambda_b+\lambda_b \lambda_a, \qquad
[\lambda_a,\lambda_b]=\lambda_a \lambda_b-\lambda_b \lambda_a.\]

As for the case of a qubit, the properties (i.) and  (ii.) of the
qudit's density  matrix are already taken into account in the
decomposition (\ref{eq:qdit}). The non-negativity requirement (iii.)
imposes further, more subtle than (\ref{eq:qball}), restrictions. A
complete characterization of  qudit's Bloch vector space,
$\mathbf{B}(\mathbb{R}^{d^2-1})$, in an arbitrary dimension, is an
open problem. However, some general properties of this space is
already known. Particularly, it can be shown that
$\mathbf{B}(\mathbb{R}^{d^2-1})$ is a convex subset of a
${d^2-1}$-dimensional unit ball
\begin{equation} \nn 
\boldsymbol{\xi}^2\leq 1\,.
\end{equation}
It being know that all pure states are concentrated on its surface.
More precisely, qudit's pure states are determined by the equation
\begin{equation} \nn 
\boldsymbol{\xi}^2 =1\,, \qquad
\boldsymbol{\xi}\vee\boldsymbol{\xi}=\boldsymbol{\xi}\,,
\end{equation}
where
\[
 (\boldsymbol{\xi}\vee\boldsymbol{\xi})_k:=
 \sqrt{\frac{d(d-1)}{2}}\,\frac{1}{d-2}\,d_{ijk}\xi_i\xi_j\,.
\]

\subsection{Composite states }

From the standpoint of quantum information theory it is mostly interesting to consider
states composed of several qubits. According to the
quantum theory axiom on composite systems ~\cite{Nielsen}, the space
of states of system, which is obtained by joining two systems $A$
and $B\,,$ represents a subspace of the tensor product of their
individual Hilbert spaces $\mathcal{H}_A$ and $\mathcal{H}_B\,$:
\begin{equation}\label{eq:compHil}
    \mathcal{H}\subset \mathcal{H}_A \otimes \mathcal{H}_B\,.
\end{equation}

The definition (\ref{eq:compHil}) in conjunction with the
superposition  principle is a source of an existence of correlations
in the joint system, which do not have any classical analog. If a
mixed state $\varrho$, describing the joint $A + B\,$ system, admit
(not necessary in unique way) representation of the form
\begin{equation}\label{eq:sep}
    \varrho=\sum_{j=1}^M \omega_j\, \varrho^A_j \otimes \varrho^B_j,\qquad
    \omega_j>0\,,
    \qquad \sum_{j=1}^M \omega_j=1\,,
\end{equation}
where  $\varrho^{A}_{j}$ and $\varrho^{B}_{j}\,$  are  density
matrices of subsystems,  then  this joint state is called
\textit{separable} \cite{Werner}. For such a state correlations
between subsystems are classically conceivable. But the states
(\ref{eq:sep}) are far to exhaust  all possible states of combined
system. The states that can not be written as (\ref{eq:sep}), are
called \textit{entangled}.

For a pair of $r$ and $s$-qudits it is useful to represent the
density matrix in the so-called Fano form~\cite{Fano57,Fano83}:
\begin{equation}\label{RhoBiPartite}
    \varrho= \frac{1}{rs} \left(\mathbb{I}_{rs}
        +\sum_{i=1}^{r^2-1} a_i\, \lambda_i \otimes \mathbb{I}_{s}
        +\sum_{i=1}^{s^2-1} b_i\,\mathbb{I}_{r} \otimes  \tau_i
        +\sum_{i=1}^{r^2-1}\sum_{j=1}^{s^2-1}
        c_{ij}\lambda_i \otimes\tau_j\right)\,.
\end{equation}
In  (\ref{RhoBiPartite}) matrices  $\lambda_i$ and $\tau_i$ are
basis elements of the $\mathrm{su}(r)$ and  $\mathrm{su}(s)$
algebras respectively. The real $(r^2-1)\times(s^2-1)$ matrix $C=
||c_{ij}||$ is so-called \textit{``correlation matrix''}.
Meaning of parameters $\boldsymbol{a}=( a_1, \ldots, a_{r^2-1})$ and
$\boldsymbol{b}=( b_1, \ldots, b_{s^2-1})$ becomes clear after
performing the partial trace operation~\cite{Vedral}:
\begin{equation} \nn 
\varrho^{(A)}:=\mt_B(\varrho)=\frac{1}{r}(\mathbb{I}_r
    +\boldsymbol{a}\cdot\boldsymbol{\lambda})\,, \qquad
\varrho^{(B)}:=\mt_A(\varrho)=\frac{1}{s}(\mathbb{I}_s
    +\boldsymbol{b}\cdot\boldsymbol{\tau}).
\end{equation}
The vectors $\boldsymbol{a}$ a $\boldsymbol{b}$ are Bloch vectors
for subsystems whose states are describing by matrices
$\varrho^{(A)}$ and $\varrho^{(B)}$, respectively.

The entanglement properties of density matrices (\ref{RhoBiPartite})
as well as more general multipartite systems admit formulation in
terms of invariants of the so-called local groups
\cite{LindenPopescu}. In the next paragraph the corresponding
notions will be introduced.

\section{The entanglement space }

\subsection{The local invariance }

 On the space of density matrices of $n$-level system the group SU(n)
acts in  adjoint manner
\begin{equation} \label{eq:untran}
    \varrho  \quad \to \quad \varrho^\prime = U^\dag \varrho\, U\,.
\end{equation}
If a quantum system is obtained by combining  $r$-subsystems with
 $n_1, n_2, \ldots , n_r\, $ levels,  the non-local properties of
 the resulting composite system  can be put into a
 correspondence with a certain decomposition of
 the unitary operations in (\ref{eq:untran}).
Namely, from all unitary actions we separate the group of so-called
\textit{local unitary transformations} (LUT)
\begin{equation}  \label{eq:LUT}
\mathrm{SU(n_1)}\otimes\mathrm{SU(n_2)}\otimes\cdots\otimes
\mathrm{SU(n_r)}\,,
\end{equation}
acting independently on the density matrices of each subsystems
\begin{equation}  \nn 
    \varrho^{(n_i)}  \quad \to \quad {\varrho^{(n_i)}}^\prime =
    {g}^\dag \varrho^{(n_i)}\, g\,
    \qquad g\in \mathrm{SU(n_i)}\,, \quad i=1,2,\ldots, r\,.
\end{equation}
Two states of composite system connected by the LUT transformations
(\ref{eq:LUT}) have the same non-local properties. The latter  can
be changed only by the rest of the unitary actions
\begin{equation} \nn 
\frac{\mathrm{SU(n)}}{\mathrm{SU(n_1)}\otimes\mathrm{SU(n_2)}\otimes\cdots\otimes
\mathrm{SU(n_r)}}\,,
\qquad
n=\prod_{i=1}^r n_i\,,
\end{equation}
generating the class of non-local transformations.

As it was mentioned in the Introduction the action of LUT is not
transitive. The equivalence of states regarding the action
(\ref{eq:LUT}) gives rise  to a decomposition of the space of
matrices into the equivalence classes (orbits). The union of these
classes, i.e., the orbit space, is customary to call as the
``entanglement space'' $\mathcal{E}_n\,.$

\subsection{Orbit space and local polynomial invariants}
\label{sec:ProgrLUT}

The main motivation for studying of $\mathcal{E}_n\,$ is necessity
to  work out qualitative criteria  and quantitative measures for
characterization of non-locality in composite systems~\cite{BengtssonZyczkowski,KusZyczkowski}.

As it was mentioned above, a canonical method for description  of the
orbit space $\mathcal{E}_n\,$ is the theory of
classical invariants~\cite{PopovVinberg}. Within this approach, starting from
the works by Linden and Popescu~\cite{LindenPopescu}, series of interesting results,
which clarify the mathematical contents of the entanglement
phenomenon, has been obtained. A considerable  progress was achieved
for pure states. As an example, we refer here to the construction
of Hilbert series for a multipartite  systems of
qubits~\cite{LuqueThibon} and classification of pure entangled
states based on the theory of hyperdeterminants~\cite{Miyake}.

Analysis of the orbit space for systems in mixed states is much more
vague. The generic questions of construction of basis for rings of
local invariants for mixed states have been considered in
~\cite{GrasslRottelerBeth,KingWelshJarvis}. With this aim the
algorithmic methods of computer algebra were
used~\cite{Sturmfels},~\cite{DerksenKemper}.\footnote{Unfortunately,
applications of the existing algorithmic methods, including the
Gr\"obner bases technique, to the analysis of  the ring of
polynomial invariants  for multipartite systems is not effective due
to the sharp growth of the number of algebraic operations with the increasing number
of qubits. }

According to the theory of invariants~\cite{PopovVinberg}, the ring
of polynomial invariants $\mathbb{C}[V]^G$,  of linear space  $V$ over
the complex numbers $\mathbb{C}$, under the action of a group $G\,,$
represents  the graded algebra
\[
\mathbb{C}[V]^G=\bigoplus_{k=1}^\infty A_k,
\]
where $A_k$ is the space of homogeneous G-invariant polynomials of
degree $k$.

The special unitary groups  $\mathrm{SU(n)}$ belong to the reductive
algebraic groups. Their ring is finitely
generated~\cite{PopovVinberg}, and $\mathbb{C}[V]^G$ is
Cohen-Macaulay type~\cite{HochsterRoberts}. However, straightforward
application of this construction to the problems of quantum
entanglement is complicated by the fact  that the space $\mathfrak{P}_+\,$,  on
which the local group (\ref{eq:LUT}) acts is not a linear space.
As it was already emphasized
in the Introduction, density matrices are the positive semi-definite
and therefore  the space of representations $\mathfrak{P}_+\,$  is nonlinear
semi-definite algebraic manifold. Bellow we suggest the trick how to
overcome this difficulty,  exemplifying  the problem in details for a system of
pair of qubits.

Let start with the construction of the  ring
 $\mathbb{C}[\,{\mathcal{H}_{4\times4}}\,]^{\mathrm{SU(2)\otimes
SU(2)}}$ of the adjoint action invariants on the space of $4\times4$
Hermitian matrices $\mathcal{H}_{4\times4}$. In order to define the
ring $\mathbb{C}[\,\mathfrak{P}_+\,]^{\mathrm{SU(2)\otimes SU(2)}}$,
note that the space of positive-definite matrices $\mathfrak{P}_+$
is subspace of $\mathcal{H}_{4\times4}\,, $ which is invariant under
the action of $\mathrm{SU(4)}\,.$ As we demonstrate below the subset
$\mathfrak{P}_+$ admits representation via the set of polynomial
inequalities \footnote{One can find a description of $\mathfrak{P}_+\,,$ similar to
the given here,  in
\cite{Kimura,ByrdKhaneja,KryszewskiZachcial}.}
\begin{equation}\label{eq:inset}
    P_a(\mathfrak{C}_2, \mathfrak{C}_3, \mathfrak{C}_4) \geq 0\,,
    \qquad a=1,2,3\,
\end{equation}
in three invariants, $\mathfrak{C}_2, \mathfrak{C}_3 $ and $
\mathfrak{C}_4\,,$ of the enveloping algebra of $\mathrm{SU(4)}$
group. From the other side, since $\mathfrak{C}_2, \mathfrak{C}_3,
\mathfrak{C}_4$ are at the same time invariants of
${\mathrm{SU(2)\otimes SU(2)}}\,$, then it is possible to construct
 in
$\mathbb{C}[\,{\mathcal{H}_{4\times4}}\,]^{\mathrm{SU(2)\otimes
SU(2)}}$ such a basis that includes these invariants. As result, having this
basis  and taking into account the inequalities (\ref{eq:inset}), we
will be able to characterize the ring
$\mathbb{C}[\,\mathfrak{P}_+\,]^{\mathrm{SU(2)\otimes SU(2)}}$
completely. According to the consideration given in the subsequent
paragraphs  a basis of the ring can be chosen in a way that only
the primary  invariants of degree  2, 3 and one secondary invariant of
degree 4 presented in the ring's  Hironaka decomposition
\cite{PopovVinberg} are constrained by the above polynomial inequalities
(\ref{eq:inset}).

\section{Non-negativity of density matrix}

To succeed in our program of construction of an optimal homogeneous
basis for the ring
$\mathbb{C}[\,\mathfrak{P}_+\,]^{\mathrm{SU(2)\otimes SU(2)}}$ let
us start with the discussion of positive semi-definiteness of
density matrices. Below the requirement of non-negativity will be formulated
 in the form of inequalities in invariants of the adjoint action
of $\mathrm{SU(n)}$ group  on
$\mathfrak{P}_+\,.$

\subsection{$\mathfrak{P}_+$ in terms of Casimirs of
$\mathrm{SU(n)}$}

A Hermitian operator is positive semi-definite if and only if all
its characteristic numbers are non-negative. The condition of
non-negativity of Hermitian operator can be
formulated solely in terms of coefficients of its characteristic
equation:
\begin{equation}\label{eq:chareq}
    |\mathbb{I}_n\,x-\varrho|
    =
    x^n-S_1x^{n-1}+S_2x^{n-2}-\ldots+(-1)^n S_n
    =0\,.
\end{equation}
The coefficients  $S_k$ in (\ref{eq:chareq}) are given as
the sums of principal minors of $k$-th order:
\[
    S_k
    =
    \sum\limits_{1\leq i_1<\ldots<i_k\leq n}
    \varrho\left(
\begin{array}{ccc}
    i_1&\ldots&i_k\\
    i_1&\ldots&i_k
\end{array}
\right)\,, \qquad k=1,\ldots,n.
\]

Since the matrix  $\varrho$ is Hermitian all its characteristic
numbers $x_k\,$ are real roots  of
the characteristic equations (\ref{eq:chareq}). When  $x_k\,$ are positive
then all $S_k$ being the  symmetric polynomials
\[S_k=\sum_{1\leq i_1\leq\ldots\leq i_k\leq n}\prod_{j=1}^k x_{i_j}\,,
\]
are non-negative real numbers.
The inverse statement is true as well;  the non-negativity of the
coefficients $S_k$ provides the non-negativity  of roots $x_k\,.$
The Descartes theorem~\cite{Vinberg}:  a number of positive roots
(taking into account their multiplicity) equals to the number of
signs changes in the sequence of the coefficients of the polynomial
equation, gives proof of this observation (see e.g.~\cite{Kimura}).

So, non-negativity of density matrices can be written in the
invariant way  as condition of non-negativity of the coefficients of
its characteristic equation:
\begin{equation}\label{eq:nonnegatcoef}
S_k \geq 0 \,, \qquad k=1,\ldots,n.
\end{equation}

We give here, for further use, the explicit form of a few first
coefficients $S_k,$ written in terms of $n$-dimensional Bloch
vector $\boldsymbol{\xi}$ \cite{Kimura,ByrdKhaneja}:
\begin{align}
    &S_2=\frac{1}{2!}\frac{n-1}{n}\,(1-\boldsymbol{\xi}\cdot\boldsymbol{\xi}),\nn\\
    &S_3=\frac{1}{3!}\frac{(n-1)(n-2)}{n^2}\,
        (1-3\,\boldsymbol{\xi}\cdot\boldsymbol{\xi}+
        2\,(\boldsymbol{\xi}\vee\boldsymbol{\xi})\cdot\boldsymbol{\xi}),\nn\\
    &S_4=\frac{1}{4!}\frac{(n-1)(n-2)(n-3)}{n^3}\,
        (1-6\,\boldsymbol{\xi}\cdot\boldsymbol{\xi}+8\,(\boldsymbol{\xi}\vee
       \boldsymbol{\xi})\cdot\boldsymbol{\xi}
    \nonumber\\
    &\qquad +3\,\frac{n-1}{n-3}(\boldsymbol{\xi}\cdot\boldsymbol{\xi})^2
    -6\,\frac{n-2}{n-3}(\boldsymbol{\xi}\vee\boldsymbol{\xi})
    \cdot(\boldsymbol{\xi}\vee\boldsymbol{\xi})).\nn
\end{align}

Besides from the restrictions (\ref{eq:nonnegatcoef}) there are
upper bounds on $S_k$ due to the normalization condition
$\mathrm{Tr}(\varrho) =1\,,\ $ $\mathrm{Tr}(\varrho^k)\leq 1\,,$ for
$k\geq 2\,.$ Note,  that the equality fulfils  for  pure states and
the maximal values of $S_k$ are achieved for the equal eigenvalues
$x_i$ of density matrices.

Finally, the positive semi-definiteness and normalizability
conditions for density matrices of $n$-level system can be written
as the following set of inequalities
\begin{equation}\label{eq:IneqSyst}
    0 \leq \,
    \frac{k!\,n^{k-1}\,S_k}{(n-1)(n-2)\ldots(n-k+1)}\,
     \leq 1\,, \quad  \qquad k=2,\ldots, n.
\end{equation}

Coefficients $S_k,\; k=1,\ldots,n$ of the characteristic equation
are invariants under the adjoint action of $\mathrm{SU(n)}\,$ group.
They are algebraically independent  and can be represented via
polynomials in the Casimir operators  of the corresponding
enveloping algebra. Below, the case $n=4\,,$ related to the system
of 2-qubits,  is considered in details and inequalities
(\ref{eq:IneqSyst}) are rewritten directly in terms of the
Casimir operators of the enveloping $\mathrm{su(4)}\,$ algebra.

\subsection{Restrictions on invariants of  {SU(4)} }

The group  SU(4) has three  Casimir operators whose expressions in
terms of the components of 15-dimensional Bloch vector
$\boldsymbol{\xi}$ (see the decomposition (\ref{eq:qdit})) can be
written as
\begin{align}
 & \mathfrak{C}_2=\boldsymbol{\xi}\cdot\boldsymbol{\xi}\,,\label{C2Casimir}\\
 & \mathfrak{C}_3=\boldsymbol{\xi}\vee\boldsymbol{\xi}\cdot\boldsymbol{\xi}\,,
 \label{C3Casimir}\\
 & \mathfrak{C}_4=\boldsymbol{\xi}\vee\boldsymbol{\xi}\cdot
 \boldsymbol{\xi}\vee\boldsymbol{\xi}\,.
 \label{C4Casimir}
\end{align}
Because for an arbitrary 4-level system the  coefficients  $S_2, S_3$ and $S_4$  of characteristic
equation of density matrix  are expressible via these  Casimir operators
\begin{align}
  &S_2=\frac{3}{8}(1-\mathfrak{C}_2)\,, \nonumber\\
  &S_3=\frac{1}{16}(1-3\mathfrak{C}_2+2\mathfrak{C}_3)\,, \nonumber\\
  &S_4=
  \frac{1}{256}((1-3\mathfrak{C}_2)^2 +8\mathfrak{C}_3-12\mathfrak{C}_4)\,, \nonumber
\end{align}
the set (\ref{eq:IneqSyst}) reduces to the following constraints on
$\mathrm{SU(4)}$ invariants
\begin{equation}\label{eq:IneqSyst4}
    \begin{array}{c}
  0\leq \mathfrak{C}_2 \leq 1\, ,\\
  0\leq 3\mathfrak{C}_2-2\mathfrak{C}_3 \leq 1\, , \\
  0\leq (1-3\mathfrak{C}_2)^2 +8\mathfrak{C}_3-12\mathfrak{C}_4 \leq 1\, .
\end{array}
\end{equation}

\begin{figure}
\begin{center}
  \includegraphics[width=8cm]{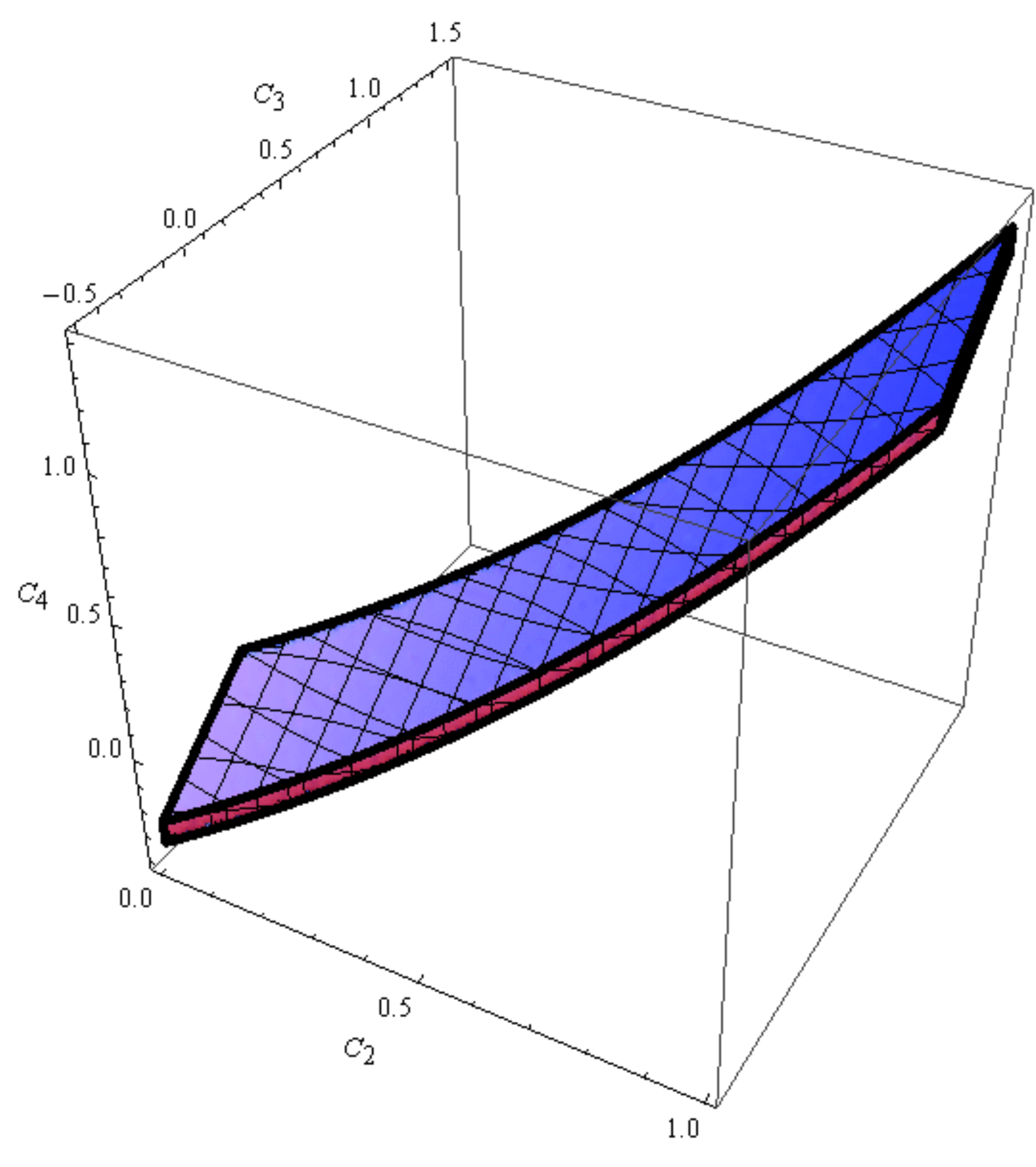}
  \caption{The allowed region for the values of Casimirs
$\mathfrak{C}_2, \mathfrak{C}_3$ and $\mathfrak{C}_4\,.$
  }\label{fig}
\end{center}
\end{figure}

In the space spanned  by invariants $\mathfrak{C}_2, \mathfrak{C}_3$ and
$\mathfrak{C}_4\,$ inequalities  (\ref{eq:IneqSyst4}) define the
bounded domain  depicted on the Figure \ref{fig}.

\section{The ring of local invariants
$\mathbb{C}[\,\mathfrak{P}_+\,]^{\mathrm{SU(2)\otimes SU(2)}}$
}\label{PolynomialInvariantRing}

Consider the density matrix of two qubits parameterized in the Fano
form:
\begin{equation}\label{eq:rho2qb}
\varrho = \frac{1}{4}\,\left[\,\mathbb{I}_2\otimes\mathbb{I}_2
 +\boldsymbol{a}\cdot \boldsymbol{\sigma}\otimes\mathbb{I}_2
 +\mathbb{I}_2\otimes\boldsymbol{b}\cdot \boldsymbol{\sigma}
 +c_{ij}\,\sigma_i\otimes\sigma_j\,\right]\,,
\end{equation}
where 3-component vectors $\boldsymbol{a}= (a_1, a_2, a_3)\,$ and $
\boldsymbol{b}=(b_1, b_2, b_3)\,$ are Bloch vectors of constituent
qubits, $\sigma_i\,, \ i =1,2,3$ - the Pauli matrices making up the
basis of algebra $\mathrm{su}(2)$:
\begin{equation}\label{PauliMatrices}
    \sigma_1 =\left(%
\begin{array}{cc}
  0 & 1 \\
  1 & 0 \\
\end{array}
\right)\,, \quad
\sigma_2 =\left(%
\begin{array}{cc}
  0 & -i \\
  i & 0 \\
\end{array}
\right)\,, \quad
\sigma_3 =\left(%
\begin{array}{cc}
  1 & 0 \\
  0 & -1 \\
\end{array}
\right)\,.
\end{equation}
The correlation matrix  $\mathrm{C}$ of pair of qubits has 9
elements $c_{ij}\,\ i,j = 1,2,3\,.$

At first, following our  program of the construction of the ring of
invariants $\mathbb{C}[\,\mathfrak{P}\,]^{\mathrm{SU(2)\otimes
SU(2)}}\,$ outlined in paragraph  \ref{sec:ProgrLUT}, we identify the space of
parameters $\boldsymbol{a}\,, \boldsymbol{b}\,$ and $\mathrm{C}\,$
with $\mathbb{R}^{15}\,,$ ignoring for a moment all restrictions due
to the non-negativity of density matrices. Apart from
this we linearize the adjoint action
 (\ref{eq:untran}) of the local group
${\mathrm{SU(2)\otimes SU(2)}}$
\begin{equation}\label{eq:linLUT}
    V_A\  \to \ V_A^\prime = L_{AB} V_B\,\qquad A,B=1,\ldots, 15\,,
\end{equation}
with  $15\times 15$ matrix $L \in \mathrm{SU(2)\otimes
SU(2)}\otimes\overline{\mathrm{{SU(2)}\otimes{SU(2)}}}$.

So, our preliminary issue is to build up the ring of polynomial
invariants of linear action of $\mathrm{SU(2)\otimes
SU(2)}\otimes\overline{\mathrm{{SU(2)}\otimes{SU(2)}}}$ group on the
linear space  $\mathbb{R}^{15}\,$. Note that the linearization
(\ref{eq:linLUT}) allows to use the prompt from the Molien formula
for the generating function of invariants for  $\pi_G$
representations of a compact group $G$~\cite{DerksenKemper}:
\begin{equation} \label{eq:Molienfunc}
M(q)= \int_{G}\mathrm{d}\mu_G\,\frac{1}{\det||\mathrm{id}-
q\pi_G||}\,,
\end{equation}
where the integral is taken over the group $G$ with the Haar measure
$\mathrm{d}\mu_G\,.$

The Molien function provides information on the polynomial
invariants ring's structure. Firstly, its formal series in powers of
parameter $q$, the so-called Hilbert-Poincare  series:
\[
M(q)=\sum_{k\geq 0} {d_k} q^k\, \in \,\mathbb{Z}[q]\,,
\]
points out the dimension,  ${d_k}$, of the space of homogeneous
invariants of degree  ${k}\,$. Secondarily, being a rational
function, (\ref{eq:Molienfunc}) admits (non-uniquely) for $q < 1\,,$
the representation
\begin{equation*}
M(q)=\frac{\sum_{k=0}^r q^{\mathrm{deg} J_k}}{\Pi_{m=1}^n(1-
q^{\mathrm{deg} K_m})}\,,
\end{equation*}
From this form of the Molien function on can conclude on the number
and order of the primary  $K_i \,, \ i=1,2,\ldots,n\,, $ and
secondary $J_i \,, \ i=1,2,\ldots,r\,, $ invariants of
the Cohen-Macaulay algebra
\begin{equation} \nn 
\mathbb{C}[V]^G=\bigoplus_{k=0}^{r} \,J_k\,\mathbb{C}[K_1, K_2,
\ldots, K_n]\,.
\end{equation}

As computations show,  the Molien function for mixed states of two
qubits can be written as \cite{GrasslRottelerBeth,KingWelshJarvis}:
\begin{equation}\label{eq:Molien2qbit}
M(q)=\frac{1+q^4+q^5+{3}q^6+{2}q^7+{2}q^8+
{3}q^9+q^{10}+q^{11}+q^{15}
}{(1-q)(1-q^2)^{3}(1-q^3)^{2}(1-q^4)^{3}(1-q^6)}\,,
\end{equation}
According to the result (\ref{eq:Molien2qbit}), a basis of ring
consists from  10 primary invariants of degree
$1,2,2,2,3,3,4,4,4,6\,$ and 15 secondary invariants of degree $
4,5,6,6,6,7,7,8,8,9,9,9,10,11,15\,$.

More detailed information on invariants dependence on
coefficients of the decomposition (\ref{eq:rho2qb}) can be extracted
using the so-called method of many-parametric generating functions
\cite{DerksenKemper}. In our case the many-parametric generating
function  depends not only on  one parameter $q$, but is a function of
three arguments, $F(a,b, c)$. The contribution from variables
$\boldsymbol{a}\,, \boldsymbol{b}\,$ and $c_{ij}\,$ into the Molien
function now is taken with the weights determined  by each
independent parameter, ${a}\,, {b}\,$ and $c\,$ respectively.

It is worth to note that the generating function $F(a,b, c)$ was
found already in the middle of 70-th of the last century
\cite{Juddatal,Quesne}, in connection with so-called problem of
\textit{``missing index''} which arose within issue
 of nuclei spectrum classification. The corresponding mathematical formulation
 and solution of the problem can be found,  e.g., in \cite{Juddatal}.
Further, in our presentation, we will mainly follow the article~\cite{Quesne}.

Consider the space of all polynomials in fifteen variables
$a_i,b_i $ and $c_{ij}\,$ $i,j=1,2,3$. In virtue of the adjoint action of the
local group the space of Bloch's parameters is decomposed into the
irreducible representations of the SO(3) $\otimes$ SO(3) group. More
precisely, the variables $a_i,b_i$ and $ c_{ij}$ are transformed according
to the representations $D_1\times D_0,\ D_0\times D_1, $ and $
D_1\times D_1$ correspondingly. Since the subspace,
$P_{s,t,q}[a_i,b_i,c_{ij}]$, of homogeneous polynomials in variables
$a_i,b_i,c_{ij}$ of degree $s,t,q\,$ correspondingly,  is invariant under the action
$\mathrm{SU(2)\otimes SU(2)}\,,$  all invariants $C$ can be
classified according to their degrees of homogeneity, i.e.,
${C}^{(s\,t\,q)}$.

Consider, following the construction suggested in ~\cite{Quesne},
the set of invariants:\footnote{Below, everywhere it is assumed the
summation over all repeated indices from one to three.}
\begin{enumerate}
\item[\empty] 3 invariants of second degree
\begin{eqnarray}\label{eq:2order}
C^{(002)}=c_{ij}c_{ij}\,, \quad C^{(200)}=a_ia_i\,, \quad
C^{(020)}=b_ib_i\,,
\end{eqnarray}
\item[\empty] 2 invariants of third degree
\begin{eqnarray}\label{eq:3order}
C^{(003)}= \frac{1}{3!}\epsilon_{ijk}\epsilon_{\alpha\beta\gamma}
c_{i\alpha}c_{j\beta}c_{k\gamma}\,, \qquad C^{(111)}=a_ic_{ij}b_j\,,
\end{eqnarray}

\item[\empty] 4 invariants of fourth degree
\begin{eqnarray}\label{eq:4order}
  C^{(004)}&=&c_{i\alpha}c_{i\beta}c_{j\alpha}c_{j\beta}\,,\\
   C^{(202)}&=&a_i a_j c_{i\alpha}c_{j\alpha}\,,\\
   C^{(022)}&=&b_\alpha b_\beta c_{i\alpha}c_{i\beta}\,,\\
   C^{(112)}&=&
    \epsilon_{ijk}\epsilon_{\alpha\beta\gamma}a_i b_\alpha
    c_{j\beta}c_{k\gamma}\,,
\end{eqnarray}

\item[\empty] 1 invariant of fifth degree
\begin{eqnarray}\label{eq:5order}
&&C^{(113)}= a_i c_{i\alpha}c_{j \alpha}c_{j \beta}b_\beta\,,
\end{eqnarray}

\item[\empty] 4  invariants of sixth degree
\begin{eqnarray}\label{eq:6order}
&&C^{(123)}=
    \epsilon_{ijk}b_i c_{\alpha j} a_\alpha c_{\beta k}c_{\beta l}b_l\,,\\
&&C^{(204)}=a_ic_{i\alpha}c_{j\alpha}c_{j\beta}c_{k\beta}a_k\,,\\
 &&C^{(024)}=b_ic_{\alpha i}c_{\alpha j}c_{\beta j}c_{\beta,
 k}b_k\,,\\
&&C^{(213)}=\epsilon_{\alpha\beta\gamma}a_\alpha c_{\beta i} b_i
c_{\gamma j} c_{\delta j} a_\delta\,,
\end{eqnarray}
\item[\empty] 2  invariants of seventh degree
\begin{eqnarray}\label{eq:7order}
&&C^{(214)}=
    \epsilon_{ijk}b_i c_{\alpha j} a_\alpha c_{\beta k}c_{\beta l}c_{\gamma l}a_l\,,\\
&&C^{(124)}= \epsilon_{\alpha\beta\gamma}a_\alpha c_{\beta j} b_j
c_{\gamma k}c_{\delta k}c_{\delta l}b_l \,,
\end{eqnarray}

\item[\empty] 2 invariants of eighth degree
\begin{eqnarray}\label{eq:8order}
&&C^{(125)}=
    \epsilon_{ijk}b_i c_{\alpha j} c_{\alpha l}b_l  c_{\beta k}c_{\beta m}c_{\gamma m}
    a_\gamma\,,\\
&&C^{(215)}= \epsilon_{\alpha\beta\gamma}a_\alpha c_{\beta
i}c_{\delta i} a_\delta c_{\gamma k}c_{\varrho k}c_{\varrho l}b_l
\,,
\end{eqnarray}

\item[\empty]2 invariants of ninth degree
\begin{eqnarray}\label{eq:9order}
&&C^{(306)}=
    \epsilon_{\alpha\beta\gamma}a_\alpha c_{\beta i}c_{\delta i}a_\delta
     c_{\gamma j}c_{\varrho j}c_{\varrho k}c_{\sigma k}a_\sigma\,,\\
&&C^{(036)}=\epsilon_{ijk}b_i c_{\alpha j}c_{\alpha l}b_l
     c_{\beta k}c_{\beta m}c_{\gamma m }c_{\gamma s}b_s\,,
\end{eqnarray}
\end{enumerate}

From these invariants the basis of
$\mathbb{C}[\,\mathfrak{P}_+\,]^{\mathrm{SU(2)\otimes SU(2)}}\,$ can
be build. As the criterion of its construction we choose the
principle of usage of basis with the minimal number of elements
involved in the definition of $\mathfrak{P}_+\,.$ Having in mind
this  rule and noting that the space $\mathfrak{P}_+\,$ is defined
in terms of the  Casimir operators of $\mathrm{SU(4)} $
group~(\ref{C2Casimir})-(\ref{C4Casimir}), we expand $\mathfrak{C}_2,
\mathfrak{C}_3, \mathfrak{C}_4$ over the set of above introduced
local invariants (\ref{eq:2order})-(\ref{eq:4order}):
\begin{align}\label{eq:KQ1}
 & \mathfrak{C}_2=\frac{1}{3}\,(C^{(200)}+{C}^{(020)}+{C}^{(002)})\,,\\
 & \mathfrak{C}_3={C}^{(111)}-{C}^{(003)}\,, \\
 & \mathfrak{C}_4=\frac{1}{6}\,
 [2({C}^{(200)}{C}^{(020)}+{C}^{(202)}+{C}^{(022)}-{C}^{(112)})
+({C}^{(002)})^2-{C}^{(004)}]\,. \label{eq:KQ3}
\end{align}
From equations (\ref{eq:KQ1})-(\ref{eq:KQ3}) it follows that one can
consider the Casimir operators $\mathfrak{C}_2, \mathfrak{C}_3,
\mathfrak{C}_4$ as the basis elements instead of scalars
$C^{(002)}\,, C^{(003)}\,$ and $\,C^{(112)}\,$.

Bear in mind this observation and using the results of
\cite{KingWelshJarvis}, where the ring
$\mathbb{C}[\,\mathbb{R}^{16}\,]^{\mathrm{SU(2)\otimes SU(2)}}\,$
was described, we define the following set consisting from 10
\textit{primary invariants}, including the Casimir operators
$\mathfrak{C}_2, \mathfrak{C}_3$,
\begin{eqnarray}
  && \mbox{deg}=4\,, \qquad \ \ K_1= 1\,, \nonumber\\
  &&\mbox{deg}=2\,,  \qquad \ \ K_2=\mathfrak{C}_2\,,\  \qquad\qquad\qquad  K_3=C^{(200)}\,,
\ \   K_4=C^{(020)}\,,\nonumber\\
  &&\mbox{deg}=3\,, \qquad \ \  K_5=\mathfrak{C}_3\,, \  \qquad\qquad\qquad   K_6= C^{(111)}\,,
  \label{eq:primary} \\
&&\mbox{deg}=4\,, \qquad  \ \   K_7=C^{(004)}\,, \ \  \ \qquad
\qquad K_8= C^{(202)}\,, \ \  K_9=
C^{(022)}\,,\nonumber\\
&&\mbox{deg}=6\,, \qquad \ \ K_{10}=C^{(204)}+C^{(024)}\,,\nonumber
\end{eqnarray}
and 15 \textit{secondary invariants} including the Casimir
$\mathfrak{C}_4$
\begin{eqnarray}\label{eq:secodary}
  && \mbox{deg}=4\,, \qquad \ \ J_1= \mathfrak{C}_4\,, \nonumber\\
  &&\mbox{deg}=5\,,  \qquad \ \ J_2=C^{(113)}\,,\nonumber\\
  &&\mbox{deg}=6\,, \qquad \ \  J_3=C^{(204)}-C^{(024)}\,, \qquad J_{8}=C^{(123)}\,,
  \ \  J_{9}=C^{(213)}\,,\nonumber\\
&&\mbox{deg}=7\,, \qquad \ \  J_{10}=C^{(214)}\,,\   \  \ \ \qquad\qquad J_{11}=C^{(124)}\,,
\label{eq:secondary}\\
&&\mbox{deg}=8\,, \qquad \ \  J_{12}=C^{(215)}\,, \  \ \   \ \qquad\qquad J_{13}=C^{(125)}\,, \nonumber\\
&&\mbox{deg}=9\,, \qquad  \ \   J_{4}=J_1J_2\,, \  \ \   \
\qquad\qquad \ \  \
J_{14}=C^{(306)}\,,  \ \  J_{15}= C^{(036)}\,,\nonumber\\
&&\mbox{deg}=10\,, \qquad \ \ J_5= J_1J_3\,,\nonumber\\
&&\mbox{deg}=11\,, \qquad \ \ J_6= J_2J_3\,,\nonumber\\
&&\mbox{deg}=15\,, \qquad \ \ J_7= J_1J_2J_3\,.\nonumber
\end{eqnarray}

We conclude that the set of homogeneous invariants
(\ref{eq:primary})-(\ref{eq:secondary}) represents the basis for the
ring  $\mathbb{C}[\,\mathfrak{P}\,]^{\mathrm{SU(2)\otimes SU(2)}}$:
\begin{equation} \nn 
\mathbb{C}[\mathfrak{P}_+]^{\mathrm{SU(2)\otimes
SU(2)}}=\bigoplus_{k=0}^{15} \,J_k\,\mathbb{C}[K_1, K_2, \ldots,
K_{10}]\,,
\end{equation}
under the condition, that two primary invariants  $K_2, K_5 $ and
one secondary invariant $J_1$ satisfy  the inequalities
(\ref{eq:IneqSyst4}).

\section{Conclusion}

An essential issue of the quantum theory of information is
qualitative and quantitative  characterization of purely  quantum
correlations caused by the entanglement of quantum states. Theory of
classical invariants provides tools for studies of the corresponding
space of entanglement, i.e., the orbit space of action of the group
of local transformations on the space of states of composite
systems. For the case we are interesting in, system of two qubits in
a mixed state, the local transformations of the density matrices
form the $\mathrm{SU(2)\otimes SU(2)}\,$ group. Its  adjoint action,
on the space of the Hermitian, unit trace matrices, identified with
$\mathbb{R}^{15}\,,$ defines the principal orbit space
\[
\mathcal{O}:=\frac{\mathbb{R}^{15}}{ \mathrm{SU(2)\otimes SU(2)}}\,,
\]
with dimension
\[
\mbox{dim}\,\mathcal{O}=15-2\times3=9.
\]

However,  the orbit space defined in such a way is not the space of
entanglement $\mathcal{E}_2\,. $ Due to the non-negativity of
density matrices the space of physical states is $\mathfrak{P}_+
\subset \mathbb{R}^{15}\,.$ In the present article we suggest the
description of $\mathfrak{P}_+ \,$  based on the polynomial
inequalities in Casimir operators of the enveloping algebra
$\mathrm{su(4)}\,.$ Furthermore, we show
 how these restrictions can be effectively taken into account
 constructing  the  basis for the  ring
$\mathbb{C}[\,\mathfrak{P}_+\,]^{\mathrm{SU(2)\otimes SU(2)}}\,,$
provided for the Hironaka decomposition with only two primary
invariants of degree 2, 3 and one secondary invariant of degree 4
constrained by the polynomial inequalities (\ref{eq:IneqSyst4}).

Concluding it is important to emphasize that without the
inequalities (\ref{eq:IneqSyst4}), the usage of local invariants for
``coordinatization'' of the space of entanglement $\mathcal{E}_2$ is
not correct.  We leave for a future publications analysis of those
constraints consequences on  the geometry  of $\mathcal{E}_2 \subset
\mathcal{O}\,$.

\section*{Acknowledgments }

This work was supported in part  the  Georgian National Science
Foundation research grant GNSF/ST08/4-405 and by the Russian Foundation
for Basic Research (grant No. 10-01-00200) and by the Ministry of
Education and Science of the Russian Federation (grant No.
3810.2010.2).


\end{document}